\documentclass[twocolumn, pre, twoside, floatfix]{revtex4-1}

\usepackage{amsmath}
\usepackage{amssymb}
\usepackage{graphicx}
\usepackage{color}
\usepackage{bm}

\def\dps{\displaystyle}
\def\eps{\varepsilon}

\def\theta{\vartheta}
\def\rho{\varrho}

\begin{document}


\title{Static dielectric properties of dense ionic fluids}

\author{Grigory Zarubin}
\email{zarubin@is.mpg.de}
\author{Markus Bier}
\email{bier@is.mpg.de}
\affiliation
{
   Max-Planck-Institut f\"ur Intelligente Systeme, 
   Heisenbergstr.\ 3,
   70569 Stuttgart,
   Germany, 
   and
   Institut f\"ur Theoretische Physik IV,
   Universit\"at Stuttgart,
   Pfaffenwaldring 57,
   70569 Stuttgart,
   Germany
}

\date{28 April 2015}

\begin{abstract}
The static dielectric properties of dense ionic fluids, e.g., room temperature
ionic liquids (RTILs) and inorganic fused salts, are investigated on different
length scales by means of grandcanonical Monte Carlo simulations.
A generally applicable scheme is developed which allows one to approximately
decompose the electric susceptibility of dense ionic fluids into the 
orientation and the distortion polarization contribution.
It is shown that at long range the well-known plasma-like perfect screening
behavior occurs, which corresponds to a diverging distortion susceptibility,
whereas at short range orientation polarization dominates, which coincides with
that of a dipolar fluid of attached cation-anion pairs.
This observation suggests that the recently debated interpretation of RTILs
as dilute electrolyte solutions might not be simply a yes-no-question but it
might depend on the considered length scale.
\end{abstract}

\maketitle

\section{\label{sec:intro}Introduction}

Room-temperature ionic liquids (RTILs) have become an extensively investigated 
topic in recent years. 
RTILs are salts with melting points at about room temperature, in contrast to
inorganic salts such as $\mathrm{NaCl}$ with a melting point of 
$1081\,\text{K}$ \cite{Janz1967}. 
The first report of an RTIL is dated to 1914, when the compound ethylammonium
nitrate with a melting temperature of $285\,\text{K}$ was synthesized
\cite{Walden1914}. 
Since the 1990s interest in RTILs has grown greatly due to the discovery of a
huge number of air- and water-stable compounds (see, e.g., 
Ref.~\cite{Weingartner2008} and the references therein).

Besides the strikingly low melting point, RTILs exhibit a number of remarkable
general, i.e., material-independent, physical properties such as an extremely
low vapour pressure \cite{Zaitsau2006, Bier2010} and a high viscosity at ambient
temperature --- the lowest viscosity of RTILs at $298\,\text{K}$ observed by
now, $21\,\text{cP}$ for $\mathrm{[C_2mim][N(CN)_2]}$ \cite{Macfarlane2001},
is more than twenty times that of water. 
All these properties are related to the combination of strong electrostatic and
highly anisotropic steric forces.

An interesting feature of the RTILs for applications is the high sensitivity of
some physical properties on the choice of the cation and the anion. 
For example the viscosity of $\mathrm{[C_4mim][BF_4]}$ is only half of that of 
$\mathrm{[C_4mim][PF_6]}$ even though these RTILs differ only in the size of
the (spherical) anions \cite{Seddon2002}. 
Therefore exploration of RTILs on the molecular level is crucial, in order to
being able to ``design'' salts with prescribed properties. 
Computer simulations are frequently used as a tool, but the challenge is to 
find an adequate description of the interaction potentials. 
Such interaction potentials are complicated and depend upon many parameters 
(e.g., bond, angle and torsion parameters as well as van der Waals terms 
\cite{Hunt2006}). 
However, the aim of the present work is not to model a specific RTIL with all
its peculiarities but to investigate ionic fluids, i.e., RTILs and
fused salts, from a general point of view which considers them mostly as fluids
being composed of ions at high densities. 
Hence, since general features are expected to occur in all systems, one can
focus on the simple system of equally-sized charged hard spheres, which is 
commonly called the restricted primitive model (RPM).

As RTILs are used as solvents in chemical studies, their polarity is one of the
most important characteristics because it describes the global solvation 
capability of the solvent. 
In order to interpret surface force apparatus (SFA) measurements, it has been proposed recently
\cite{Gebbie2013_1} to view pure ionic liquids as dilute electrolyte solutions
with a few mobile ions in an effective solvent made of temporarily paired ions. 
Whereas these SFA data have been doubted \cite{Perkin2013,
Gebbie2013_2}, the
interpretation of RTILs as dilute electrolyte solutions has attracted some
interest \cite{Lee2014}. 
For non-conducting fluids the static dielectric constant $\eps$ is well defined
and precisely measurable (see, e.g., Ref.~\cite{Bezman1997,MohsenNia2013}). 
However, conducting fluids, such as ionic fluids, are well-known to perfectly 
screen external charges at \emph{long} ranges \cite{Stillinger1968, Hansen1975}
so that the corresponding static dielectric constant is infinitely large. 
On the other hand, it is also well-known that cations and anions form a 
characteristic alternating pair structure at \emph{short} ranges 
\cite{Stillinger1968, Hansen1975}, which, when considering neighboring
cation-anion pairs, is analogous to the charge separation inside dipolar 
molecules. 
Therefore, it appears that the dielectric properties of dense ionic fluids 
depend on the length scale.

In the present work the analysis is based on the response of bulk ionic fluids
to a static nonuniform electric field which spatially varies on a particular
length scale.
It has to be stressed that here the focus is on nonuniform static electric
fields, the response onto which is described in terms of the 
wavenumber-dependent dielectric function $\eps(k)$, whereas numerous 
experimental studies consider uniform time-dependent fields, which give rise to
the frequency-dependent dielectric function $\eps(\omega)$ (see, e.g., 
\cite{Weingartner2001, Weingartner2006, Daguenet2006, Lui2011}). 
Both quantities, $\eps(k)$ and $\eps(\omega)$, are not easily related because
the former describes the equilibrium structure whereas the latter quantifies 
the dynamics in uniform electric fields.

The dielectric properties of a substance can be interpreted in terms of two
well-known mechanisms \cite{Atkins1998}: Orientation polarization refers to
the rotation of dipolar moments upon keeping the magnitude constant, whereas
distortion polarization describes the change of magnitude of dipolar moments at
constant orientations.
The according orientation susceptibility $\chi_\text{ori}(k)$ and distortion
susceptibility $\chi_\text{dis}(k)$ both contribute to the total electric 
susceptibility $\chi(k)=\eps(k)-1=\chi_\text{ori}(k) + \chi_\text{dis}(k)$.
However, in order to infer the dominant polarization mechanism as a function
of the wave number $k$, one has to somehow determine the decomposition of the
observable $\chi(k)=\eps(k)-1$ into the orientation and the distortion
contribution.
Perfect screening in ionic fluids and the according divergence $\chi(k)\to
\infty$ in the limit $k\to0$ \cite{Stillinger1968, Hansen1975} corresponds to a
dominating distortion polarization in the long-wavelength limit.
This long-range behavior of ionic fluids is in sharp contrast to that of
dipolar fluids, whose electric susceptibility $\chi(k)$ attains a finite limit
$\chi(0)$ as $k\to0$.

In order to achieve the decomposition of the electric susceptibility $\chi(k)$
of an ionic fluid into the orientation and the distortion susceptibility
$\chi_\text{ori}(k)$ and $\chi_\text{dis}(k)$, respectively, the following
approach is proposed: In addition to determine the electric susceptibility
$\chi(k)$ of the ionic fluid composed of cations and anions, a corresponding
dipolar fluid is considered whose particles are overall charge-neutral
dumbbells formed by gluing together pairs of cations and anions of the ionic
fluid. 
This dipolar fluid does not exhibit distortion polarization, i.e., 
$\chi_\text{dis}(k)=0$, but only pure orientation polarization, i.e., 
$\chi(k)=\chi_\text{ori}(k)$, and the latter can be expected to be similar to
$\chi_\text{ori}(k)$ of the corresponding ionic fluid at short ranges.
The applicability of this approach is not restricted to the RPM investigated
here but it can be used for any ionic fluid model.

A brief description of the theoretical methods as well as the considered models
is given in Sec.~\ref{sec: 2}. The results of actual calculations are discussed
in Sec.~\ref{sec: 3} from which general conclusions on the dielectric properties
of ionic fluids are drawn in Sec.~\ref{sec: 4}.

\section{Models and methods \label{sec: 2}}

\subsection{Restricted primitive and dumbbell model\label{subsec:models}}

\begin{figure}[!t]
  \includegraphics[width=8cm]{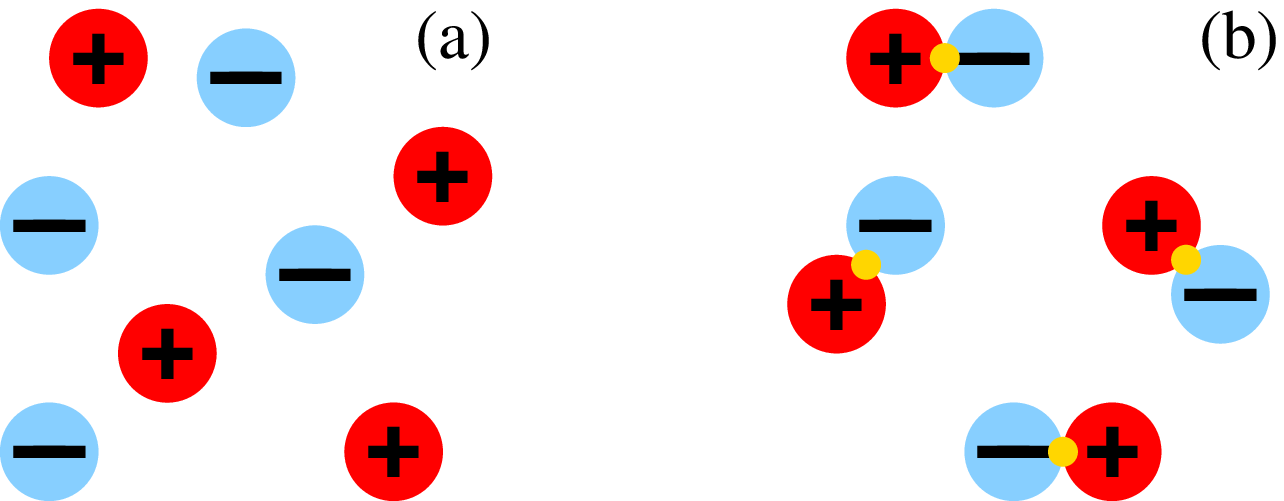}
  \caption{Sketch of (a) the restricted primitive model (RPM) and (b) the
           corresponding dumbbell model (DM) with particles being constructed
           by pairwise gluing together oppositely charged hard spheres of the
           RPM.}
  \label{fig: 1}
\end{figure}

The model used to represent the ionic fluid is the three-dimensional restricted
primitive model (RPM), i.e., a collection of $N/2$ positively and $N/2$
negatively charged hard spheres of equal diameter $\sigma$ and equal absolute
valencies $|z_+|=|z_-|$ (see Fig.~\ref{fig: 1}(a)). 
The interaction potential between two ions of species $i$ and $j$ with 
$i,j\in\{+,-\}$ at positions $\bm{r}_i$ and $\bm{r}_j$, respectively, can be 
described as
\begin{align}
  \label{eq: HSILPotential}
  \beta U_{ij}(\bm{r}_i,\bm{r}_j) = 
  \begin{cases}
      \dps\frac{z_i z_jl_\text{B}}{|\bm{r}_i - \bm{r}_j|} & 
      , |\bm{r}_i - \bm{r}_j| \geq \sigma \\
      \infty & , |\bm{r}_i - \bm{r}_j| < \sigma
   \end{cases}
\end{align}
with the vacuum Bjerrum length $l_\text{B} = \beta e^2/(4\pi \eps_0)$, where 
$e$ is the elemenatry charge, $\beta = 1/(k_\text{B} T)$ denotes the inverse 
temperature, and $\eps_0$ is the vacuum permittivity.

In this work, the dielectric properties of the RPM (see Fig.~\ref{fig: 1}(a))
are compared with those of a corresponding dipolar fluid model whose particles
are composed of one cation and one anion of the RPM glued together (see 
Fig.~\ref{fig: 1}(b)). 
The particles within this dumbbell model (DM) possess three positional and two
orientational degrees of freedom, in contrast to six translational degrees of
freedom of a pair of ions within the RPM.
Obviously, all configurations of $N/2$ dumbbells correspond to possible
configurations of the RPM with $N/2$ positive and $N/2$ negative hard spheres,
but not all configurations of the RPM can be realized within the DM.
The interaction energy between two dumbbell particles is given by the sum of
contributions Eq.~(\ref{eq: HSILPotential}) of the constituent charged hard
spheres.

\subsection{Dielectric properties}

In this work the dielectric properties of ionic and dipolar fluids are studied
by analyzing the linear response of the models introduced in the previous
Subsec.~\ref{subsec:models} in the presence of a weak static nonuniform 
external electric field. 
Within the linear response regime the dielectric properties are given by the
dielectric function tensor $\overleftrightarrow{\eps}(\bm{k})$ or, 
equivalently, by the electric susceptibility tensor 
$\overleftrightarrow{\chi}(\bm{k})=\overleftrightarrow{\eps}(\bm{k})-1$.
However, since electrostatic fields are purely longitudinal due to Faraday's
law, the polarization field is also purely longitudinal in isotropic fluids.
Therefore, only the longitudinal components (parallel to the wave vector
$\bm{k}$)
\begin{align}
   \eps(\bm{k})& := \eps_\|(\bm{k})
                  = \frac{\bm{k}\cdot\overleftrightarrow{\eps}(\bm{k})\cdot
                    \bm{k}}{\bm{k}^2}
                 \quad\text{and}\quad\notag\\
   \chi(\bm{k})& := \chi_\|(\bm{k}) 
                 = \frac{\bm{k}\cdot\overleftrightarrow{\chi}(\bm{k})\cdot
                   \bm{k}}{\bm{k}^2}
                 = \eps(\bm{k})-1
\end{align}
are of relevance here.

It can be shown \cite{Hansen2008} that the longitudinal dielectric function 
$\eps(\bm{k})$ is related to the charge-charge structure factor
\begin{align}
   S_{zz} (\bm{k}) = \frac{1}{N} 
   \left\langle \sum_m z_m \exp (-i \bm{k}\cdot\bm{r}_m) 
                \sum_n z_n \exp ( i \bm{k}\cdot\bm{r}_n) \right\rangle
   \label{eq:Szz}
\end{align}
via
\begin{equation}
\label{eq: Susceptibility_IL}
   \frac{1}{\eps(\bm{k})} = 
   1 - \frac{4\pi l_\text{B}}{\bm{k}^2}\frac{N}{V}S_{zz} (\bm{k}).
\end{equation}
In the present work the averaging in Eq.~(\ref{eq:Szz}) is obtained by means of
grandcanonical Monte Carlo simulations of the RPM and the DM.

\subsection{Grandcanonical Monte Carlo simulations\label{subsec:montecarlo}}

In order to determine the charge-charge structure factor $S_{zz}(\bm{k})$ in
Eq.~(\ref{eq:Szz}), grandcanonical Monte Carlo simulations with Metropolis
sampling \cite{Metropolis1953, Allen1987, Frenkel2002} were performed in
a cubic box of side length $V^{1/3}=10\sigma$ with periodic boundary conditions
applying Ewald's method \cite{Ewald1921, Allen1987, Frenkel2002}.
In a first step, quick simulation runs were used to determine the relation
between the chemical potential and the mean density $N/V$ for fixed temperature.
For calculation of the charge-charge structure factor $S_{zz}(\bm{k})$ 
typically $10^6 - 10^7$ relaxation and $10^7 - 10^8$ evaluation steps were used.
A Monte Carlo step involved a random selection of one of the following possible
moves:
\begin{itemize}
\item translation of an ion (RPM) or of a dumbbell (DM) inside a small 
      cube-like environment of size $d^3$ with probability $P_\text{trans}$,
\item insertion of a cation and an anion (RPM) or of a dumbbell (DM) with 
      probability $(1-P_\text{trans})/2$, or
\item removal of a cation and an anion (RPM) or of a dumbbell (DM) with 
      probability $(1-P_\text{trans})/2$.
\end{itemize}
Insertions and removals of charge-neutral entities guarantee global charge
neutrality during the whole simulation. Values of $P_\text{trans}$ and $d$ were
chosen in the way that the resulting rate of acceptance of the trial states did
not drop below $30\%$.
The Monte Carlo code has been validated by comparison to results obtained
using the molecular dynamics package ESPResSo \cite{Arnold2013, Limbach2006}.

\section{Results and discussion \label{sec: 3}}

In the present work the RPM (Sec.~\ref{subsec:models}) is considered as a 
representative ionic fluid.
Its dielectric function $\eps(\bm{k})$ is determined by using 
Eq.~(\ref{eq: Susceptibility_IL}) via calculating the charge-charge structure
factor $S_{zz}(\bm{k})$, Eq.~(\ref{eq:Szz}), by means of grandcanonical Monte 
Carlo simulations (see Subsec.~\ref{subsec:montecarlo}).
In the following discussion thermodynamic states are considered with the 
packing fractions $\eta = \pi N\sigma^3/(6V)\in [0.08,0.33]$ and the 
temperatures $T^* = \sigma/l_\text{B} \in [1/3,1]$. 
These conditions correspond to the plasma parameters $\Gamma=
2\eta^{1/3}/T^*\in [0.86,4.15]$.
Since the temperatures $T^*$ are well above the critical temperature 
$T^*_c\lesssim0.1$ of the RPM \cite{Fisher1994}, no influence of the 
vapor-liquid phase transition is expected to occur.
Moreover, $T^*$ is also well above the temperature range $T^*\lesssim1/4$,
where significant Bjerrum ion-pairing is expected to occur \cite{Fisher1994},
which renders the RPM a dipolar fluid already in the gas phase.
The high temperatures $T^*$ used here allow one to study the crossover from
plasma-like to dipolar-fluid-like behavior. The general trend of that crossover for varying $\eta$ and $T^*$ is captured by a simple approximative calculation described later.

\begin{figure}[!t]
  \includegraphics{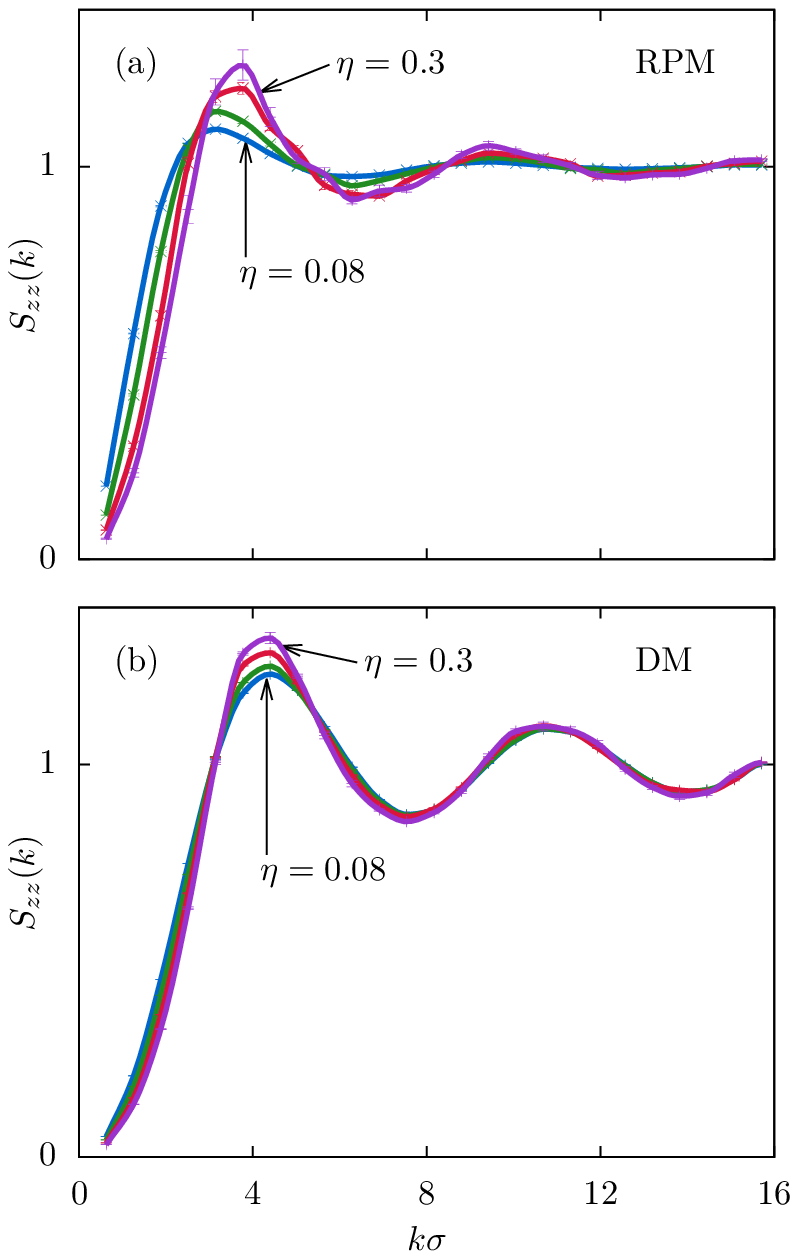}
  \caption{Charge-charge structure factor of (a) the RPM 
           (Sec.~\ref{subsec:models}) and (b) the DM (Sec.~\ref{subsec:models})
           for the temperature $T^*=1$ and various values of the packing 
           fraction $\eta\in \{0.08,0.14,0.23,0.3\}$.}
  \label{fig: 2}
\end{figure}

Figures \ref{fig: 2}(a) and (b) display the charge-charge structure factors 
$S_{zz}(\bm{k})$ of the RPM ionic fluid (Sec.~\ref{subsec:models}) and of the
DM (Sec.~\ref{subsec:models}), respectively. 
The calculation was performed several (5-10) times for each packing fraction 
separately, both for the RPM and the DM. This procedure allows one to calculate
the mean value of $S_{zz}(\bm{k})$ as well as its standard deviation shown as
error bars in Figs.~\ref{fig: 2}(a) and (b).
It is apparent that $S_{zz}(\bm{k})$ of the ionic fluid is sensitive to density
changes only for very low packing fractions $\eta$.
The slightly more pronounced oscillations of $S_{zz}(\bm{k})$ within the DM
(Fig.~\ref{fig: 2}(b)) as compared to those within the RPM 
(Fig.~\ref{fig: 2}(a)) are perhaps an artifact of the DM, within which two 
charged hard spheres are kept exactly at a distance $\sigma$, whereas the 
principal peak of the cation-anion pair distribution function within the RPM 
has a finite width \cite{Kalyuzhnyi2014}.

\begin{figure}[!t]
  \includegraphics{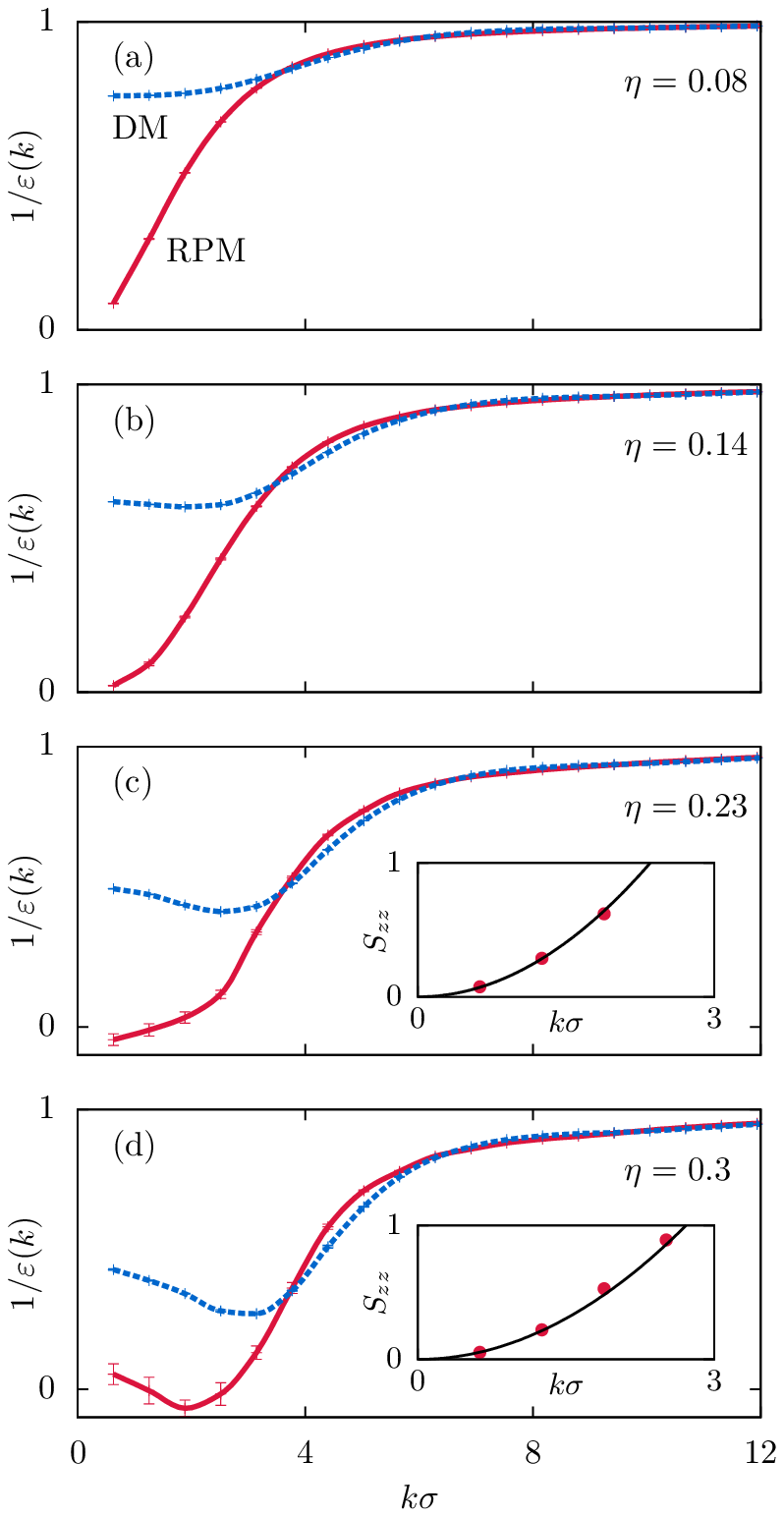}
  \caption{Inverse dielectric functions $1/\eps(\bm{k})$ of the RPM ionic fluid
           (red solid lines, see Sec.~\ref{subsec:models}) and of the DM 
           dipolar fluid (blue dashed lines, see Sec.~\ref{subsec:models}) at
           temperature $T^*=1$ and packing fractions 
           $\eta\in\{0.08,0.14,0.23,0.3\}$. 
           Whereas $1/\eps(\bm{k}\to0)$ becomes small (i.e., $\eps(\bm{k}\to0)$
           becomes large) for the ionic fluid, the dielectric function of the 
           dipolar fluid approaches a finite value $\eps(0)$ for small 
           wavenumbers. 
           The inset in panels (c) and (d) compares the charge-charge structure
           factor $S_{zz}(\bm{k})$ obtained by means of Monte Carlo simulations
           (red circles) with the asymptotic behavior Eq.~(\ref{eq:Szzasymp})
           (black solid line).}
  \label{fig: 3}
\end{figure}

Using Eq.~(\ref{eq: Susceptibility_IL}), Fig.~\ref{fig: 3} displays the inverse
dielectric functions $1/\varepsilon(\bm{k})$ of the RPM and of the DM.
The perfect screening condition \cite{Stillinger1968, Hansen2008} implies 
$1/\varepsilon (\bm{k}) \to 0$ for $\bm{k} \to 0$ for an ionic fluid 
due to the asymptotic behavior of the charge-charge structure factor
\cite{Hansen2008}
\begin{align}
   S_{zz}(\bm{k}) \stackrel{\bm{k}\to0}{\simeq} \frac{\bm{k}^2}{\kappa^2}
   \label{eq:Szzasymp}
\end{align}
with the Debye length $1/\kappa$ being given by $\kappa^2 = 
4\pi l_\text{B}N/V$, which is shown in the insets of Figs.~\ref{fig: 3}(c) and
\ref{fig: 3}(d). 
Unfortunately, the error bars of the three leftmost data points for $\eta=0.23$
(Fig.~\ref{fig: 3}(c)) and of the four leftmost data points for $\eta=0.3$ 
(Fig.~\ref{fig: 3}(d)) in the plots of the inverse dielectric function 
$1/\varepsilon (\bm{k})$ of the RPM are too large to conclusively infer the 
functional form in the long range limit $\bm{k} \to 0$ at these packing 
fractions. 
However the error bars of the RPM data in the interesting range of medium and
short separations, $|\bm{k}|\sigma \gtrsim 2\pi$, and of the DM data are small.
Since the DM can exhibit only orientation polarization, its dielectric function
$\eps(\bm{k}\to0)$ approaches a finite value of the ``dielectric constant'' 
$\eps(0)$, which increases upon increasing the packing fraction $\eta$ (see 
Fig.~\ref{fig: 3}). 

\begin{figure}[!t]
  \includegraphics{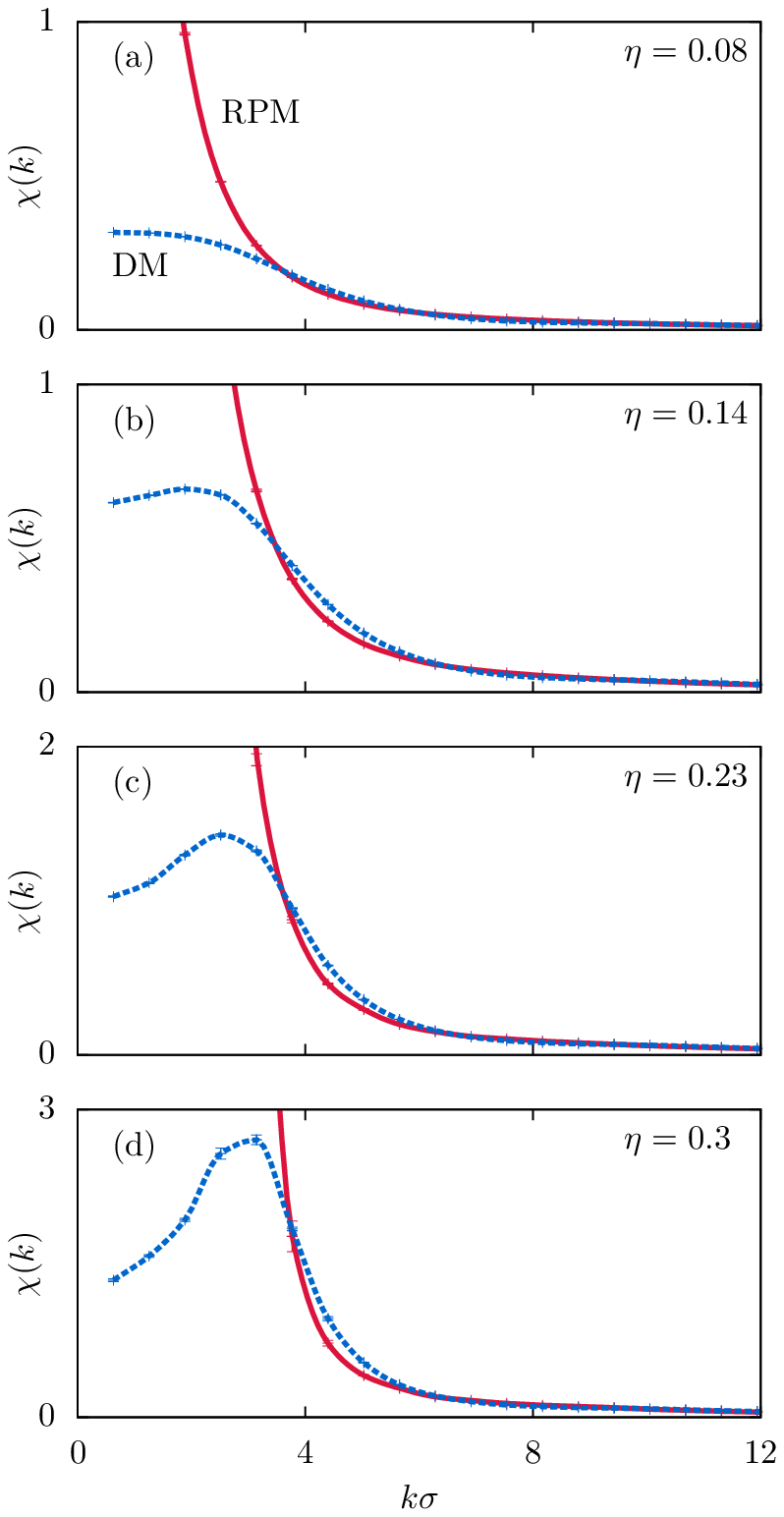}
  \caption{Electric susceptibility $\chi(\bm{k})=\eps(\bm{k})-1$ of the RPM
           ionic fluid (red solid lines, see Sec.~\ref{subsec:models}) and of
           the DM dipolar fluid (blue dashed lines, 
           see Sec.~\ref{subsec:models}) at temperature $T^*=1$ and packing
           fractions $\eta\in\{0.08,0.14,0.23,0.3\}$.
           Whereas $\chi(\bm{k}\to0)$ diverges for the ionic fluid due to the
           perfect screening of the external charges, the susceptibility of the
           dipolar fluid approaches a finite value $\chi(0)$ for small 
           wavenumbers.
           At large wavenumbers $|\bm{k}|\sigma\gtrsim 2\pi$ the electric
           susceptibility $\chi(\bm{k})=\chi_\text{ori}(\bm{k})+
           \chi_\text{dis}(\bm{k})$ of the RPM almost coincides with that of
           the DM, which, by construction, possesses only orientation 
           polarization whose electric susceptibility can be expected to be 
           close to $\chi_\text{ori}(\bm{k})$ of the RPM (see 
           Sec.~\ref{subsec:models}).}
  \label{fig: 4}
\end{figure}

\begin{figure}[!t]
  \includegraphics{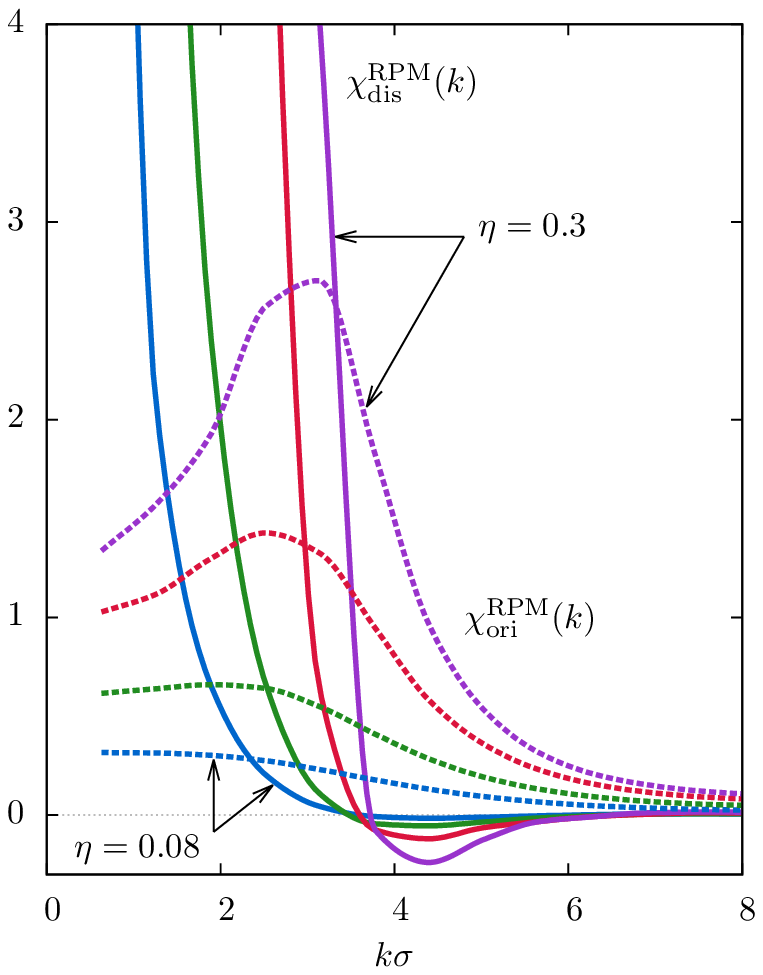}
  \caption{Distortion susceptibilities 
          $\chi_\text{dis}^\text{RPM}(\bm{k})$ (solid lines) and orientation
          susceptibilities $\chi_\text{ori}^\text{RPM}(\bm{k})$ (dashed lines)
          of the RPM at temperature $T^*=1$ for packing fractions $\eta \in 
          \{0.08,0.14,0.23,0.3\}$. 
          Perfect screening corresponds to the divergence of 
          $\chi_\text{dis}^\text{RPM}(\bm{k})$ in the long wavelength limit 
          $|\bm{k}|\to 0$.  
          The region of negative distortion susceptibility, 
          $\chi_\text{dis}^\text{RPM}(\bm{k}) < 0$, can be interpreted as
          overscreening. 
          At sufficiently large wave numbers orientation polarization dominates
          over distortion polarization with the crossover wave numbers
          increasing with the packing fraction $\eta$ (see Fig.~\ref{fig: 6}).}
  \label{fig: 5}
\end{figure}

In order to determine the dominant polarization mechanism in an ionic fluid the
inverse dielectric functions $1/\eps(\bm{k})$ of the RPM and of the DM in 
Fig.~\ref{fig: 3} are converted into the electric susceptibilities 
$\chi(\bm{k})=\eps(\bm{k})-1$.
Figure~\ref{fig: 4} shows that the electric susceptibility 
$\chi^\text{RPM}(\bm{k})=\chi_\text{ori}^\text{RPM}(\bm{k})+
\chi_\text{dis}^\text{RPM}(\bm{k})$ of the RPM, which comprises a contribution
$\chi_\text{ori}^\text{RPM}(\bm{k})$ due to orientation polarization and a 
contribution $\chi_\text{dis}^\text{RPM}(\bm{k})$ due to distortion
polarization, almost coincides in the range $|\bm{k}|\sigma\gtrsim 2\pi$ with
the electric susceptibility $\chi^\text{DM}(\bm{k})=
\chi_\text{ori}^\text{DM}(\bm{k})$ of the DM, which, by construction, exhibits
only orientation polarization with susceptibility
$\chi_\text{ori}^\text{DM}(\bm{k})$.
Since for sufficiently large packing fractions $\eta$ it can be expected that
the orientation susceptibility $\chi_\text{ori}^\text{RPM}(\bm{k})$ of the RPM
is identical to the electric susceptibility $\chi_\text{ori}^\text{DM}(\bm{k})$
of the DM, i.e., $\chi_\text{ori}^\text{RPM}(\bm{k})=
\chi_\text{ori}^\text{DM}(\bm{k})$ for all wavenumbers $\bm{k}$, one can infer
the orientation and the distortion susceptibility of the RPM separately:
\begin{align}
   \chi_\text{ori}^\text{RPM}(\bm{k}) 
   &= \chi^\text{DM}(\bm{k}), \notag\\
   \chi_\text{dis}^\text{RPM}(\bm{k}) 
   &= \chi^\text{RPM}(\bm{k})-\chi^\text{DM}(\bm{k}).
   \label{eq:decomp}
\end{align}

Figure~\ref{fig: 5} clearly indicates that, at $T^*=1$, orientation
polarization is the dominant mechanism of the RPM at sufficiently
large wave numbers $\bm{k}$, whereas distortion polarization is dominating at
sufficiently small wave numbers $\bm{k}$.
The weak distortion polarization, i.e., 
$\chi_\text{dis}^\text{RPM}(\bm{k})\approx0$ at large wave numbers $\bm{k}$ 
(see Fig.~\ref{fig: 5}) can be attributed to the 
impenetrable hard cores of the ions of the RPM. The crossover wavenumber 
between distortion-dominated and orientation-dominated polarization increases 
with packing fraction $\eta$.
There is a certain interval of wavenumbers $|\bm{k}|\sigma\approx4\dots5$ with
$\chi_\text{dis}^\text{RPM}(\bm{k})<0$ (see Fig.~\ref{fig: 5}), which
indicates overscreening, which is caused by steric effects, i.e., by the hard ion
cores, too.
The distortion susceptibility of the RPM diverges for $\bm{k} \to 0$ according
to $\chi_\text{dis}^\text{RPM}(\bm{k})\sim1/\bm{k}^2$, which corresponds to the
perfect screening property of plasmas \cite{Stillinger1968, Hansen2008}.
Hence, it has been shown here, that the RPM ionic fluid exhibits dielectric
properties similar to a dipolar fluid at short range, whereas it behaves 
plasma-like at long range.

\begin{figure}[!t]
  \includegraphics{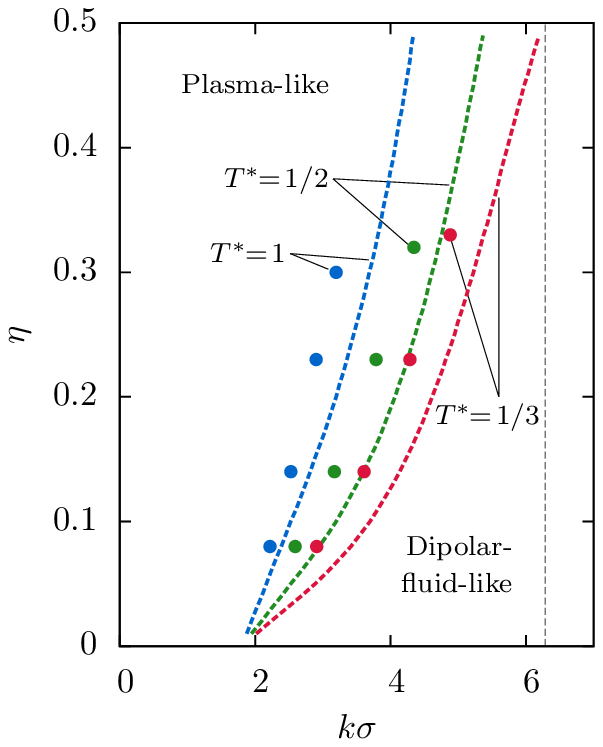}
  \caption{Crossover wave numbers $k^\times(\eta,T^*)\sigma$ of the RPM between
         plasma-like behavior for $k < k^\times(\eta,T^*)$ and 
         dipolar-fluid-like behavior for $k > k^\times(\eta,T^*)$ as functions
         of the packing fraction $\eta$ and of the temperature $T^*$.
         The crossover wave number $k^\times(\eta,T^*)$ is defined by equal
         orientation and distortion susceptibilities,
         $\chi_\text{ori}^\text{RPM}(k^\times)=
         \chi_\text{dis}^\text{RPM}(k^\times)$ (see Eq.~(\ref{eq:decomp})).
         The dots ($\bullet$) correspond to simulation data whereas the dashed
         lines are obtained by means of the simple approximative calculation
         described in the main text.
         The latter captures the correct trends of increasing values
         $k^\times(\eta,T^*)\sigma$ upon increasing the packing fraction 
         $\eta$ or decreasing the temperature $T^*$.
         The thin vertical dashed line marks $k\sigma=2\pi$, which corresponds
         to an upper bound of $k^\times(\eta,T^*)\sigma$.}
\label{fig: 6}
\end{figure}

Figure~\ref{fig: 6} displays the crossover wave numbers $k^\times(\eta,T^*)
\sigma$ (dots $\bullet$), defined by equal orientation and distortion 
susceptibilities, $\chi_\text{ori}^\text{RPM}(k^\times)=
\chi_\text{dis}^\text{RPM}(k^\times)$, as functions of the packing fraction 
$\eta$ and of the temperature $T^*$.
The RPM exhibits plasma-like behavior for $k < k^\times(\eta,T^*)$ and 
dipolar-fluid-like behavior for $k > k^\times(\eta,T^*)$.
The general trend is that of increasing values of $k^\times(\eta,T^*)\sigma$
upon increasing the packing fraction $\eta$ or decreasing the temperature 
$T^*$.
However, it can be expected that $k^\times(\eta,T^*)<2\pi/\sigma$ for any set
of parameters $(\eta,T^*)$ since an external electric field oscillating with a
wave length equal to the ion diameter $\sigma$ (see the thin vertical dashed
line in Fig.~\ref{fig: 6}) cannot lead to distortion polarization, i.e., the
crossover $\chi_\text{ori}^\text{RPM}(k^\times)=\chi_\text{dis}^\text{RPM}
(k^\times)$ has to occur at some smaller wave number.
The dashed lines in Fig.~\ref{fig: 6} correspond to an approximation of
$k^\times(\eta,T^*)\sigma$ with $\chi^\text{RPM}(\bm{k})$ in 
Eq.~(\ref{eq:decomp}) being approximated within the mean spherical 
approximation (MSA) of the RPM \cite{Hansen2008} and with 
$\chi^\text{DM}(\bm{k})$ in Eq.~(\ref{eq:decomp}) being approximated by
the expression of a single dumbbell particle in an external electric field.
Whereas this simple approximation slightly overestimates the value of 
$k^\times(\eta,T^*)\sigma$, the general trends of increasing values of 
$k^\times(\eta,T^*)\sigma$ upon increasing the packing fraction $\eta$ or 
decreasing the temperature $T^*$ are captured correctly.
Hence for dense ionic fluids, e.g., inorganic fused salts ($\eta\approx0.5,
T^*\approx1/30$) or RTILs ($\eta\approx0.5,T^*\approx1/50$), one can expect
plasma-like behavior in a very wide range of wave numbers $k<k^\times$ with
$2\pi/k^\times$ corresponding almost to the size of the ions.

\section{Conclusions \label{sec: 4}}

The main observation of the previous Sec.~\ref{sec: 3}, that the RPM ionic 
fluids exhibits dielectric properties similar to a dipolar fluid at short range
whereas it behaves plasma-like at long range, does not hinge on any 
peculiar property of the RPM and can hence be expected to be made for other
ionic fluids, too.
Moreover, our approach in Sec.~\ref{sec: 3} to decompose the electric 
susceptibility $\chi(\bm{k})=\chi_\text{ori}(\bm{k})+\chi_\text{dis}(\bm{k})$
into a contribution $\chi_\text{ori}(\bm{k})$ due to orientation polarization
and a contribution $\chi_\text{dis}(\bm{k})$ due to distortion polarization by
introducing a corresponding dipolar fluid made of cation-anion compounds applies
to the case of other ionic fluids, too.
In general, the structure of the cation-anion compounds comprising the dipolar
fluid corresponding to an ionic fluid can be conjectured on the basis of, e.g.,
the pair distribution function, which is routinely calculated for numerous
ionic fluid models (see, e.g., Ref.~\cite{Padua2007,Dommert2014}).
Hence, the method described in Sec.~\ref{sec: 3} could be a useful tool to
assess the bulk dielectric properties of general ionic fluid models.

The results of Sec.~\ref{sec: 3} are restricted to bulk ionic fluids and they
cannot be applied quantitatively in the context of the discussion on the 
interpretation of SFA measurements in an RTIL environment
\cite{Gebbie2013_1,
Perkin2013, Gebbie2013_2}, which is related to confined ionic fluids.
The reason for this restriction is that the static dielectric function for
non-uniform systems is of the form $\eps(\bm{k},\bm{k'})$ due to the absence
of translational symmetry.
However, the qualitative picture occurring in Sec.~\ref{sec: 3} suggests, that
strongly confined ionic fluids tend to behave as dipolar fluids whereas they
progressively exhibit plasma-like properties upon relaxing the confinement.
This suggests that the recently debated interpretation of RTILs as dilute 
electrolyte solutions \cite{Gebbie2013_1, Lee2014} might not be simply a 
yes-no-question but it might depend on the considered length scale.

The finding in Sec.~\ref{sec: 3}, that the static dielectric properties of
ionic fluids depend on the length scale, may be considered complementary to
the observation that the quantification of the polarity of RTILs depends 
on the intrinsic time scale of the measurement \cite{Lui2011}.

In summary, based on grandcanonical Monte Carlo simulations of the restricted
primitive model and a corresponding dumbbell model, it has been argued that
bulk ionic fluids at small length scales are expected to behave as dipolar 
fluids, for which orientation polarization dominates, whereas at large length
scales plasma-like behavior occurs, for which distortion polarization dominates.
In the style of Ref.~\cite{Lui2011}, one can conclude that the static 
dielectric properties of ionic fluids depend on the length scale on which they
are looked at.


\end{document}